# Exploiting Process Variations to Secure Photonic NoC Architectures from Snooping Attacks

Sai Vineel Reddy Chittamuru, Member, IEEE, Ishan G Thakkar, Member, IEEE,
Sudeep Pasricha, Senior Member, IEEE,
Sairam Sri Vatsavai, Student Member, IEEE, Varun Bhat, Member, IEEE,

*Abstract*— The compact size and high wavelength-selectivity of microring resonators (MRs) enable photonic networks-on-chip (PNoCs) to utilize dense-wavelength-division-multiplexing (DWDM) in their photonic waveguides, and as a result, attain high bandwidth on-chip data transfers. Unfortunately, a Hardware *Trojan* in a PNoC can manipulate the electrical driving circuit of its MRs to cause the MRs to snoop data from the neighboring wavelength channels in a shared photonic waveguide, which introduces a serious security threat. This paper presents a framework that utilizes process variation-based authentication signatures along with architecture-level enhancements to protect against data-snooping Hardware *Trojans* during unicast as well as multicast transfers in PNoCs. Evaluation results indicate that our framework can improve hardware security across various PNoC architectures with minimal overheads of up to 14.2% in average latency and of up to 14.6% in energy-delay-product (EDP).

## I. INTRODUCTION

To cope with the growing performance demands of modern Big Data and cloud computing applications, the complexity of hardware in modern chip-multiprocessors (CMPs) has increased steadily. To reduce the hardware design time of these complex CMPs, third-party hardware IPs (3PIP) are frequently used. Typically, a 3PIP vendor provides a soft IP for the CMP design, which is then sent to a foundry for fabrication. An adversary in the utilized electronic design automation (EDA) tool provider can modify the synthesized mask data and the resultant hardware implementation of the soft IP during fabrication, which can introduce security risks [1] [2]. For instance, the presence of Hardware Trojans (HTs) in the final hardware implementation can lead to leakage of sensitive information from modern CMPs [3]. Thus, security researchers that have traditionally focused on software-level security are now increasingly interested in overcoming hardware-level security risks.

This research is supported by grants from University of Kentucky and NSF (CCF-1813370).
*Sai Vineel Reddy Chittamuru* is with Micron Technology, Inc, Austin, TX. E-mail: schittamuru@micron.com. *Ishan G Thakkar* is with the Electrical and Computer Engineering Department, University of Kentucky, Lexington, KY 40506. E-mail: igthakkar@uky.edu. *Sudeep Pasricha* is with the Electrical and Computer Engineering Department, Colorado State University, Fort Collins, CO 80523. E-mail: sudeep@colostate.edu. *Sairam Sri Vatsavai* is pursuing M.S in Electrical Engineering at University of Kentucky, Lexington, KY 40506. E-mail: ssr226@uky.edu. *Varun Bhat* is with Qualcomm, San Diego, CA. E-mail: varunbhat.kn@gmail.com.

Many CMPs today use electrical networks-on-chip (ENoCs) for inter-core communication. ENoCs use packet-switched network fabrics and routers to transfer data between on-chip components [4]. Recent developments in silicon photonics have enabled the integration of photonic components and interconnects with CMOS circuits on a chip. Photonic NoCs (PNoCs) provide several prolific advantages over their metallic counterparts (i.e., ENoCs), including the ability to communicate at near light speed, larger bandwidth density, and lower dynamic power dissipation [5]. These advantages motivate the use of PNoCs for inter-core communication in modern CMPs [6].

Several PNoC architectures have been proposed to date (e.g., [7], [9]). These architectures employ on-chip photonic links, each of which connects two or more gateway interfaces. A gateway interface (GI) connects the PNoC to a cluster of processing cores. Each photonic link comprises one or more photonic waveguides and each waveguide can support a large number of dense-wavelength-division-multiplexed (DWDM) wavelengths. A wavelength serves as a data signal carrier. Typically, multiple data signals are generated at a source GI in the electrical domain (as sequences of logical 1 and 0 voltage levels) which are modulated onto the multiple DWDM carrier wavelengths simultaneously, using a bank of modulator MRs at the source GI [10]. The data-modulated carrier wavelengths traverse a link to a destination GI, where an array of detector MRs filter them and drop them on photodetectors to regenerate electrical data signals.

In most architectures, each GI in a PNoC is able to send and receive data in the optical domain on multiple (often all) utilized carrier wavelengths [9]. Therefore, each GI has a bank of modulator MRs (i.e., modulator bank) and a bank of detector MRs (i.e., detector bank). Each MR in a bank resonates with and operates on a specific carrier wavelength. In this manner, the excellent wavelength selectivity of MRs and DWDM capability of waveguides are utilized to enable high bandwidth parallel data transfers in PNoCs.

Similar to CMPs with ENoCs, the CMPs with PNoCs are expected to use several 3PIPs, and therefore, are vulnerable to security risks [11]. For instance, if the entire PNoC used within a CMP is a 3PIP, then a hardware implementation of this PNoC in a potentially malicious foundry can have HTs introduced within the control units of the PNoC's GIs. These HTs can snoop on packets in the network. These packets can be transferred to a malicious core (a core running a malicious application) in the CMP to extract sensitive information.

Unfortunately, MRs are especially susceptible to security threatening manipulations from HTs. In particular, the MR tuning circuits that are essential for supporting data broadcasts/multicasts and to counteract MR resonance shifts due to process variations (PV) make it easy for HTs to retune MRs and initiate snooping attacks. To enable data broadcast/multicast in PNoCs, the tuning circuits of detector MRs partially detune them from their resonance wavelengths [8], [12]-[13] (described in more detail in Section 3), such that a significant portion of the photonic signal energy in the data-carrying wavelengths continues to propagate in the waveguide to be absorbed in the subsequent detector MRs. On the other hand, process variations (PV) cause resonance wavelength shifts in MRs [14]. Techniques to counteract PV-induced resonance shifts in MRs involve retuning the resonance wavelengths by using carrier injection/depletion or thermal tuning [6], implemented through MR tuning circuits. An HT in the GI can manipulate the abovementioned tuning circuits of detector MRs to partially tune the detector MR to a passing wavelength in the waveguide, which enables snooping of the data that is modulated on the passing wavelength. Such covert data snooping is a serious security risk in PNoCs.

In this work, we present a framework that protects data from snooping attacks and improves hardware security in PNoC architectures. Our framework has low overhead and is easily implementable in any existing DWDM-based PNoC architecture without major changes to the architecture. Our novel contributions are:

1) We analyze security risks in photonic devices and extend this analysis to the link level, to determine the impact of these risks on PNoC architectures;
2) We propose a circuit-level scheme called Privy Data Encipherment Scheme (*PDES*), which integrates two strategies to protect data from snooping HTs: (i) exploitation of MRs' PV profiles to generate unclonable signatures/keys for data encryption, and (ii) privy incorporation of communication metadata in the data encryption-decryption mechanism to protect the identity of utilized encryption keys;
3) We propose an architecture-level Reservation-Assisted Metadata Protection Scheme (*RAMPS*) that conceals the communication metadata from snooping HTs to further enhance data security in DWDM-based PNoCs;
4) We combine these circuit-level and architecture-level schemes (*PDES* and *RAMPS*) into a holistic framework called SOTERIA; and analyze it on the Firefly [8], SwiftNoC [35], and LumiNoC [13] PNoC architectures.

## II. RELATED WORK

Several prior works, e.g., [11], [16], [17], [36]-[41] discuss the presence of security threats in ENoCs and have proposed solutions to mitigate them. In [11], a three-layer security system approach was presented by using data scrambling, packet certification, and node obfuscation to enable protection against data snooping attacks. A symmetric-key based cryptography design was presented in [16] for securing the NoC. In [17], a framework was presented to use permanent keys and temporary session keys for NoC transfers between secure and non-secure cores. Three different mechanisms to protect hybrid circuit-packet switched ENoC routers from timing channel attacks were presented in [36]. In [37], a detailed security analysis related to planned obsolescence of TSV-based 3D NoCs was presented. A non-interference based adaptive routing scheme to secure NoCs from side channel and Denial-of-Service (DoS) attacks was proposed in [38]. In [39], a packet validation technique was proposed to protect compromised network-on-chip (NoC) architectures from fault injection side channel attacks and covert HT communications. A security enhanced NoC was proposed in [40], which is able to identify traffic anomalies and handle distributed timing attacks. In [41], a self-contained Network-on-Chip (NoC) firewall at the network interface (NI) layer was presented, which, by checking the physical address against a set of rules, rejects untrusted CPU requests to the on-chip memory, and thereby protects all legitimate processes running in a multicore SoC. A*ll of these prior works focus on security enhancement in ENoCs, but none of them has analyzed security risks in photonic devices and links; or considered the impact of these risks on PNoCs.*

Fabrication-induced PV impact the cross-section, i.e., width and height, of photonic devices, such as MRs and waveguides. In MRs, PV causes resonance wavelength drifts, which can also be caused due to thermal variations (TV) [58]. These PV+TV induced resonance drifts in MRs can be counteracted by using device-level techniques such as thermal tuning or current injection tuning [6]. Current injection tuning can induce blue shifts in the resonance wavelengths of MRs using carrier injection into MRs, whereas thermal tuning can induce red shifts in MR resonances through heating of MRs using integrated heaters. To remedy PV, the use of device-level tuning techniques is inevitable; but their use also enables partial detuning of MRs that can be used to snoop data from a shared photonic waveguide. Prior works, such as [57], propose link-level techniques to address the PV issue. In [57], the assignments of MRs to carrier wavelengths are rearranged to minimize the power consumption of compensating the PV induced resonance shifts. In addition, prior works [18], [19], [29] discuss the impact of PV-remedial techniques on crosstalk noise and propose techniques to mitigate it. The impact of PV-remedial techniques on crosstalk was quantified in [18]. In [19], an encoding mechanism based on the PV-profile of MRs was proposed to mitigate crosstalk noise caused by PV-remedial techniques. A double MR based crosstalk mitigation strategy was proposed in [29] to reduce crosstalk noise. However, *none of these prior works analyzes the impact of PV-remedial techniques on hardware security in PNoCs.*

Our proposed framework in this paper enables security against snooping attacks during not only unicast communications but also multicast communications in PNoCs. Our framework is network agnostic, mitigates PV, and has minimal overhead, while improving security for any DWDM-based PNoC architecture.

## III. HARDWARE SECURITY CONCERNS IN PNoCs

### A. Threat Model

Modern chip-multiprocessor (CMP) designs often incorporate intellectual properties (IPs) (e.g., reusable units of logic, functionality, or layout) from third party providers. Use of such third-party IPs (3PIPs) reduces the complexity and time required for putting together soft CMP IPs (e.g., RTL level source codes), which are then synthesized into fabrication mask data using EDA tools and PDKs. These synthesized mask data are then sent to the external foundry for fabrication. The common practice of employing third party EDA tools for synthesis and foundries for fabrication partially relinquishes the designers' control over the final CMP hardware implementations. As a result, it becomes possible for an adversary in the EDA tool provider to implant Hardware Trojans (HTs) into the synthesized mask data and final CMP hardware implementations. These HTs are very small malicious alterations in complex IPs that are activated only under certain conditions to introduce security threats such as Denial of Service (DoS) and Data Snooping.

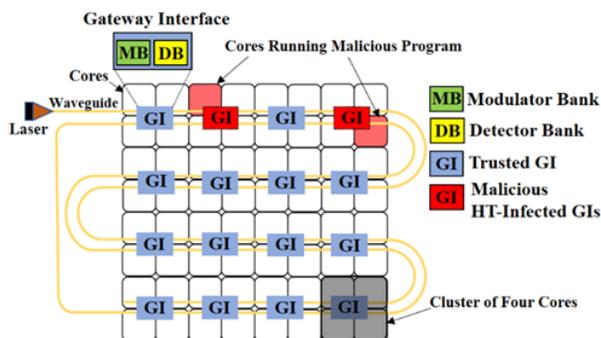

Fig. 1: Schematic of a compromised PNoC with its processing cores running malicious programs and gateway interfaces (GIs) infected by Hardware Trojans (HTs).

Unfortunately, PNoCs are also expected to incorporate the use of 3PIPs and fabrication through third party foundry in their hardware design cycle, which exposes them to security threats related to HTs. Fig. 1 shows the schematic layout of a typical PNoC, in which the gateway interfaces (GIs) are connected with each other through photonic waveguides in a serpentine topology. As mentioned earlier, a GI connects a cluster of cores (e.g., four cores here) to the PNoC, and each GI has at least one bank of modulators to enable photonic data transmission and at least one bank of detectors to enable photonic data reception. Each of these modulators and detectors employs control circuits to enable its active operation and redressal from PV induced resonance shifts (further discussed in section III.B). An adversary in the foundry can introduce HTs in these control circuits. It can also partner with software providers to introduce malicious application programs to be run on the final CMP hardware. As shown in Fig. 1, there can be HTs in the control circuits of multiple GIs, as well as instances of malicious programs simultaneously running on multiple cores. These HT-infected GIs can partner with malicious program instances to create security threats in the PNoC.

### B. Device-Level Security Concerns

Process variation (PV) induced undesirable changes in MR widths and heights cause "shifts" in MR resonance wavelengths, which prevents accurate modulation (at the source) or filtering for detection (at the receiver) with MRs. This shift can be remedied by using current injection tuning and thermal tuning. The current injection tuning method injects (or depletes) free carriers into (or from) the Si core of an MR using the electrical tuning circuit, which reduces (or increases) the MR's refractive index owing to the electro-optic effect, thereby remedying the PV-induced red (or blue) shift in the MR's resonance wavelength. In contrast, a practical way of using thermal tuning for remedying PV induced resonance shifts is to produce blue-shift fabrication bias in MRs so that the resonance wavelengths of the MRs at fabrication are smaller than (towards the blue end of the spectrum from) their desired operating wavelengths [48]. Doing so automatically converts any PV induced red or blue shift in an MR resonance into a net blue-shift, given that a large enough blue-shift fabrication bias is applied to the MR to begin with [48]. This net blue-shift then can be compensated for by heating up the MR to a corresponding higher temperature using the integrated micro-heater, which can be controlled using a similar electrical tuning circuit. Current injection tuning can provide a tuning range of only 1.5nm at most [42], but it incurs relatively low latency and power overheads [43]. In contrast, thermal tuning incurs high latency and power overheads, but it can provide a larger tuning range of about 6.6nm [43]. Moreover, thermal tuning can cause "thermal crosstalk" to change the temperature of adjacent channel waveguides, causing them to operate under a temperature gradient, and reducing the reliability of light coupling between the waveguide and MRs. Therefore, although the use of either current injection tuning or thermal tuning is theoretically sufficient for remedying the PV induced resonance shifts, we envision a tuning mechanism as described in [43] that can combine the benefits of both these techniques by intelligently using the least power consuming technique from current injection tuning and thermal tuning to lock the MR resonance with the nearest available carrier wavelength either towards the blue end or the red end of the spectrum. Using the tuning mechanism from [43], the resonance of no MR in the PNoC needs to be tuned for more than half the channel gap. In addition, the electro-optic effect (i.e., carrier injection/depletion) produced by the same electrical tuning circuit as used for current injection tuning is typically used to enable modulator MRs to move in and out of resonance (i.e., switch ON/OFF) with the utilized carrier wavelengths for signal modulation [7]. An HT can manipulate these electrical tuning circuits that are used for PV remedy (current injection tuning + thermal tuning) and signal modulation, which may lead to malicious operation of modulator or detector MRs, as discussed next.

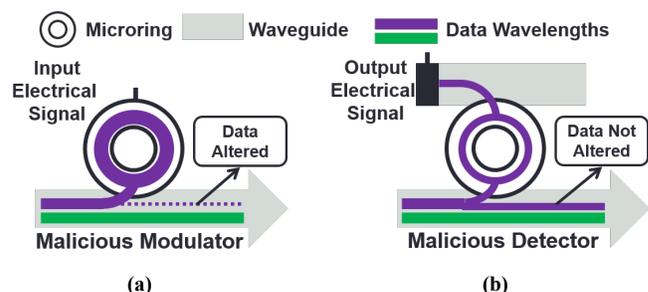

Fig. 2: Data transfer in a DWDM-based photonic waveguide, with (a) a malicious modulator MR leading to data corruption, and (b) a malicious detector MR leading to data snooping.

Fig. 2(a) shows the malicious operation of a modulator MR.

A malicious modulator MR is partially tuned to a data-carrying wavelength (shown in purple) that is passing by in the waveguide. The malicious modulator MR draws some power from the data-carrying wavelength, which can ultimately lead to data corruption as optical '1's in the data can lose significant power to be altered into '0's. Moreover, a malicious modulator MR can also cause denial of data communication by fully tuning to a data-carrying wavelength and completely drawing all its power from the waveguide (not shown in Fig. 2). Alternatively, a malicious detector (Fig. 2(b)) can be partially tuned to a passing data-carrying wavelength, to filter only a small amount of its power and drop it on a photodetector for data duplication. This small amount of filtered power does not alter the data in the waveguide so that it continues to travel to its target detector for legitimate communication [12]. Further, a malicious detector MR can also cause data corruption (by partially tuning to a wavelength) and denial of communication (by fully tuning to a wavelength). Thus, both malicious modulator and detector MRs can corrupt data (which can be detected and corrected) or cause Denial of Service (DoS) type of security attacks. In addition, malicious detector MRs can also snoop data from the waveguide without altering it. Such covert snooping attacks present a security threat in photonic links.

*C. Link-Level Security Concerns*

Typically, a photonic link is comprised of one or more DWDM-based photonic waveguides. A DWDM-based photonic waveguide uses a modulator bank (a series of modulator MRs) at the source GI and a detector bank (a series of detector MRs) at the destination GI. DWDM-based waveguides can be broadly classified into four types: single-writer-single-reader (SWSR), single-writer-multiple-reader (SWMR), multiple-writer-single-reader (MWSR), and multiple-writer-multiple-reader (MWMR). As SWSR, SWMR, and MWSR waveguides are subsets of an MWMR waveguide we restrict our link-level analysis to MWMR waveguides only.

An MWMR waveguide typically passes through multiple GIs, connecting the modulator banks of some GIs to the detector banks of the remaining GIs. Thus, in an MWMR waveguide, multiple GIs (referred to as source GIs) can send data using their modulator banks and multiple GIs (referred to as destination GIs) can receive (read) data using their detector banks. Fig. 3 presents an example MWMR waveguide with two source GIs and two destination GIs. Fig. 3(a) and 3(b), respectively, present data corruption by a malicious source GI and data snooping by a malicious destination GI, on this MWMR waveguide. In Fig. 3(a), the modulator bank of source GI $S_1$ is sending data to the detector bank of destination GI $D_2$. When source GI $S_2$, which is in the communication path, becomes malicious with an HT in its control logic, it can manipulate its modular bank to modify the existing '1's in the data to '0's. This ultimately leads to data corruption. For example, in Fig. 3(a), $S_1$ is supposed to send '0110' to $D_2$, but because of data corruption by malicious GI $S_2$, '0010' is received by $D_2$.

Let us consider another scenario for the same data communication path (i.e., from $S_1$ to $D_2$). When destination GI $D_1$, which is in the communication path, becomes malicious with an HT in its control logic, the detector bank of $D_1$ can be partially tuned to the utilized wavelength channels to snoop data. In the example shown in Fig. 2(b), $D_1$ snoops '0110' from the wavelength channels that are destined to $D_2$. The snooped data from $D_1$ can be transferred to a malicious core within the CMP to determine sensitive information.

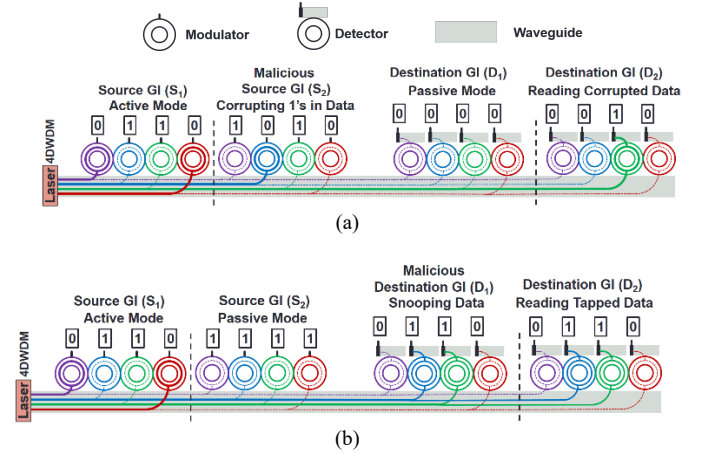

Fig. 3: Impact of (a) malicious modulator (source) bank, (b) malicious detector (destination) bank on data in DWDM-based photonic waveguides.

In addition to data corruption and snooping attacks (illustrated in Fig. 3), malicious MRs in MWMR links can also cause Denial of Service (DoS) attacks as briefly described in Section III.*B*. The adverse impacts of such DoS attacks can be mitigated by implementing cross-layer strategies that promote self-monitoring and self-adaptation in PNoCs (e.g., [45] presents such strategies for a hybrid electrical-wireless NoC). On the other hand, the impacts of data corruption attacks can be mitigated by employing conventional error detection and correction mechanisms (e.g., [46]). However, our focus in this work is to prevent data snooping attacks. This is because snooping attacks from malicious destination GIs are hard to detect, as they do not disrupt the intended communication among CMP cores. Moreover, it is difficult to detect a snooping attack by monitoring the drop in the received signal power level due to the partially extracted optical power by the snooping MRs. This is because thermal-induced fluctuations can also cause similar drop in the received signal power level, which makes it almost impossible to infer whether a snooping attack or thermal-induced power fluctuation is the real cause behind the drop in the received power level. Therefore, there is a pressing need to address the security risks imposed by snooping GIs in DWDM-based PNoC architectures. To address this need, we propose a novel framework called *SOTERIA* that improves hardware security in DWDM-based PNoC architectures.

IV. SOTERIA FRAMEWORK: OVERVIEW

Our proposed multi-layer *SOTERIA* framework enables secure unicast and multicast communications in DWDM-based PNoC architectures. Fig. 4 gives a high-level overview of this framework. From the figure, the framework integrates two security enhancing strategies: (i) Privy Data Encipherment Scheme (*PDES*), and (ii) Reservation-Assisted Metadata Protection Scheme (*RAMPS*). *PDES* uses the PV profiles of the detector MRs to generate unclonable keys that are used for encrypting data before transmission. To enable secure sharing of the generated encryption keys, *PDES* incorporates communication metadata (e.g., identity of target destination GI, type of communication – unicast or multicast) into its data encryption-decryption process. *PDES* is sufficient to protect data from snooping GIs as long as the utilized communication metadata can be kept secret. But the security of communication metadata from snooping attacks is at risk in many PNoC architectures (e.g., [11], [27]) as these PNoCs use the same waveguide to transmit both the communication metadata and actual data. To further enhance security for these PNoCs, we devise an architecture-level Reservation-Assisted Metadata Protection Scheme (*RAMPS*) that uses a secure reservation waveguide to avoid the stealing of communication metadata by snooping GIs. The combination of *PDES* and *RAMPS* schemes makes it very difficult for HTs and malicious program instances to snoop data. The next two sections present details of our *PDES* and *RAMPS* schemes.

Note that our encryption-based data security method, which is part of our proposed SOTERIA framework, is different from the traditional Authenticated Encryption (AE) [55][56] used for electronic NoCs in three ways. First, our framework proposes to utilize the information on the PV profiles of MRs as the system entropy to seed the symmetric key generation process. Doing so can achieve better security, as it can make the initial conditions, and hence, the final outcome of the key generation process much more difficult for an attacker to predict. Second, our PDES scheme, which is part of our SOTERIA framework, generates and utilizes destination specific encryption keys. Therefore, only the source nodes that have the specific key corresponding to a destination node can send encrypted data packets to the destination node, thereby ensuring the authenticity of data packets in addition to their confidentiality. Third, the traditional AE scheme for electronic NoCs does not protect the confidentiality of metadata itself, which is very important to achieve dependable routing of data packets to intended destinations. In contrast, our proposed RAMPS scheme can protect the confidentiality of metadata as well. Thus, our SOTERIA framework (PDES + RAMPS) utilizes encryption and decryption processes for PNoCs in a novel and functionally different manner compared to the traditional AE mechanism for electronic NoCs.

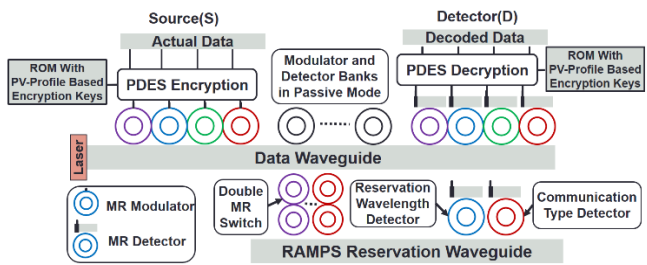

Fig. 4: Overview of proposed *SOTERIA* framework that integrates a circuit-level Privy Data Encipherment Scheme (*PDES*) and an architecture-level Reservation-Assisted Metadata Protection Scheme (*RAMPS*).

## V. Privy Data Encipherment Scheme (PDES)

As discussed earlier (Section III.C), malicious destination GIs can snoop data from a shared waveguide. One way of addressing this security concern is to use data encryption so that the malicious destination GIs cannot decipher the snooped data. For the encrypted data to be truly undecipherable, the malicious GIs in a PNoC must not be able to clone the algorithm (or method) used to generate the keys used for data encryption. To generate unclonable encryption keys, our Privy Data Encipherment Scheme (*PDES*) uses the PV profiles of MRs. As discussed in [14], PV induces random shifts in the resonance wavelengths of the MRs used in a PNoC. These resonance shifts can be in the range from -3nm to 3nm [14]. The MRs that belong to different GIs in a PNoC have different PV profiles. In fact, the MRs that belong to different MR banks of the same GI also have different PV profiles. Due to their random nature, these MR PV profiles cannot be cloned by the malicious GIs, which makes the encryption keys generated using these PV profiles truly unclonable. Using the PV profiles of MRs, *PDES* can generate a unique encryption key for each MR bank in a PNoC.

For the encrypted data to be truly undecipherable, the unclonable key used for data encryption should be kept secret from the snooping GIs, which can be challenging as the identity of the snooping GIs in a PNoC is not known. Therefore, it becomes very difficult to decide whether or not to share the encryption key with a destination GI (that can be malicious) for data decryption. Moreover, in case of multicast communication with a PNoC, an encrypted data message can be communicated to more than one destination GI simultaneously. This makes it even more difficult to decide which destination GIs to share the encryption key with. To resolve this key-sharing conundrum, *PDES* makes use of the following information about the communicated data messages: *(i)* identity of target destination GIs, *(ii)* type of data communication – unicast or multicast. This important information about the communicated data messages is referred to as communication metadata henceforth. *PDES* employs this communication metadata for its key generation, key sharing, and data encryption-decryption processes, which are described next.

### A. Key Generation and Sharing with PDES

As discussed earlier, *PDES* utilizes the PV profiles of MRs to generate unclonable encryption keys. But MRs employed in a PNoC can be categorized as modulator MRs, detector MRs, and switches (i.e., MRs that can route photonic wavelength signals in PNoCs). This raises an obvious question: PV profiles of which category of MRs should be used to generate unclonable encryption keys? Generally, the PV profiles of MRs of any category can be utilized to generate unclonable encryption keys. But *PDES* utilizes PV profiles of the destination GIs' detector MRs to generate encryption keys, as doing so renders the following useful properties to the generated keys: *(i)* different encryption keys can be generated for unicast and multicast communications, *(ii)* a unique encryption key can be generated for each destination GI of every MWMR waveguide, and *(iii)* the generated encryption keys can be shared and used securely, yielding an efficient solution to the key-sharing conundrum. We will discuss the origins and benefits of these properties later in this section.

Another question that begs attention is: how to measure and put the PV profiles of thousands of MRs a PNoC might have, to work for key generation? *PDES* generates encryption keys from MR PV profiles during the testing phase of the CMP. For that, it measures the PV-induced resonance shifts in all MRs of every destination GI's detector bank in situ, using dithering signal-based control circuits from [15]. For each detector MR of every destination GI, the corresponding control circuit (i.e., from [15]) generates an anti-symmetric analog error signal. This analog error signal is proportional to the PV-induced resonance shift in the detector MR during the testing phase of the CMP, whereas it to proportional to the net resonance shift in the detector MR due to the combined effects of PV and dynamic thermal variations (TV) during the dynamic operation of the CMP. During the dynamic operation of the CMP, *PDES* uses this anti-symmetric analog error signal to control the carrier injection into and heating of the MR to remedy the induced net shift in its resonance (due to the combined effects of PV and TV). On the other hand, during the static testing phase of the CMP, *PDES* converts the analog error signal from each individual detector MR into an 8-bit digital error signal. Thus, an 8-bit digital error signal is generated for each detector MR of every destination GI. We consider 64 DWDM wavelengths per waveguide, and hence, we have 64 detector MRs in every destination GI's detector bank. As an 8-bit digital error signal is generated from each individual detector MR, a total of 64 digital error signals (8-bits each) are generated for each detector bank of every destination GI. These 64 digital error signals (of 8-bit each) per detector bank per destination GI are utilized in two different ways to generate two different encryption keys: one key for unicast communications and one key for multicast communications, as described next.

**Unicast communications:** For each destination GI, our *PDES* scheme appends all 64 digital error signals (of 8-bits each) corresponding to the 64 detector MRs in a randomly selected permutational order to create a unique 512-bit encryption key for unicast communications (referred to as *unicast key* henceforth). Different destination GIs append their respective 64 8-bit error signals in different permutational orders (randomly selected at design time) to derive their *unicast* keys. Having a uniquely different *unicast key* for every destination GI enables *PDES* to protect each *unicast key* from malicious snooping GIs. This is because it eliminates the need of sharing a *unicast key* that is specific to a destination GI with any other secure or malicious destination GI. Thus, our *PDES* scheme enables low cost generation and secure utilization of PV-based encryption keys to be used for unicast communications. Since we use unique keys for each destination GI, the frequency of the same key being used is low, thus making this scheme reasonably immune to various ciphertext attacks (e.g., [47]) typically used for deducing the key. Moreover, because of a very large number of available permutational orders (i.e., 64!) in which the constituent 64 error signals can be appended to derive a *unicast key*, it becomes highly unlikely to have the encryption keys of any two destination GIs in the PNoC to be the same, even if the available PV range for the PNoC does not scale up with the number of GIs in the PNoC. Thus, *PDES* is highly scalable.

**Multicast communications:** In PNoCs, some messages (e.g., cache coherence messages) need to be communicated with multiple destination GIs simultaneously, which establishes the need for multicast communication capability in PNoCs. As discussed earlier, to enable data multicast in a waveguide that connects multiple destination GIs, the tuning circuit of each target destination GI partially detunes its detector MRs from their resonance wavelengths [8], [12]-[13], such that a significant portion of the photonic signal energy in the data-carrying wavelengths continues to propagate in the waveguide to be absorbed in all subsequent target destination GIs. Thus, the partial tunability of the destination GIs' detector MRs enables efficient multicasting in PNoCs. Like unicast communications, *PDES* uses data encryption to secure multicast communications as well. But the key used for encrypting such multicast messages (referred to as *multicast key*, henceforth) cannot be designed to be specific to one or more destination GIs. This is because each multicast message is generally sent to a group of destination GIs (referred to as *multicast group*). Many different combinations of destination GIs are possible to form many possible target *multicast groups*. Therefore, utilizing a specific key for each target *multicast group* can incur excessive overheads for key generation, storage, and sharing. To avoid this overhead, a single *multicast key* should be used that is common for all possible target multicast groups, such that any target multicast group can decrypt its received multicast message using the *multicast key*. Such *multicast keys* cannot be protected from malicious destination GIs, as they may very likely be shared with all destination GIs, including the malicious ones, for data decryption. However, this does not mean that multicast communications cannot be secured using data encryption. *PDES* employs communication metadata (i.e., the information about the type of communication – unicast or multicast) in its data encryption process to achieve secure multicast communications. To generate a *multicast key* for an MWMR waveguide, *PDES* XORs all 512-bit *unicast keys* corresponding to all destination GIs that share the MWMR waveguide. Thus, *PDES* generates one 512-bit *multicast key* for every MWMR waveguide in a PNoC.

In summary, our use of such indirectly and in situ measured (using the dithering signal based sensing/control mechanism from [15]) process variation profile of MRs for encryption key generation provides three-pronged advantages: (1) It enables generation of keys that are different from one another not only across different detector banks and gateway interfaces of a single chip, but also across different chips, which can strengthen the capabilities of ensuring the confidentiality, authenticity and integrity of intra-chip as well as inter-chip communications. (2) It enables the generation of destination-specific unclonable keys in situ, deeming the use of external random key generators unnecessary. This integrated in-situ approach can reduce significant time and effort during mass production of CMPs. (3) This integrated in-situ approach also enables easy upgrades of the key generation process in the future; not only the process variation profile, but also the thermal variation and aging profiles of MRs can be indirectly measured and leveraged down the road using the same sensing mechanism from [15] to generate unclonable keys.

### B. Data Encryption-Decryption with PDES

To understand how the 512-bit *unicast* and *multicast keys* are utilized to encrypt data in photonic links, consider Fig. 5 which depicts an example photonic link that has one MWMR waveguide and connects the modulator banks of two source GIs ($S_1$ and $S_2$) with the detector banks of three destination GIs ($D_1$, $D_2$, and $D_3$). As there are three destination GIs on this link, *PDES* creates three 512-bit *unicast keys* corresponding to them and stores them at respective destination GIs and both source GIs. Moreover, *PDES* creates one 512-bit *multicast key* specific to the MWMR waveguide and stores it at all source and destination GIs (i.e., $S_1$, $S_2$, $D_1$, $D_2$, and $D_3$). Thus, in Fig. 4, every source GI stores three 512-bit *unicast keys* (for destination GIs $D_1$, $D_2$, and $D_3$) and the 512-bit *multicast key* in its local ROM, whereas every destination GI stores only its corresponding 512-bit *unicast key* along with the 512-bit *multicast key* in its ROM. When data is to be transmitted by a source GI, depending on the type of communication, the appropriate key from its local ROM is used to encrypt data at the packet-level granularity, using the XOR cipher algorithm [50] that performs an XOR between the key and the 512-bit data packet.

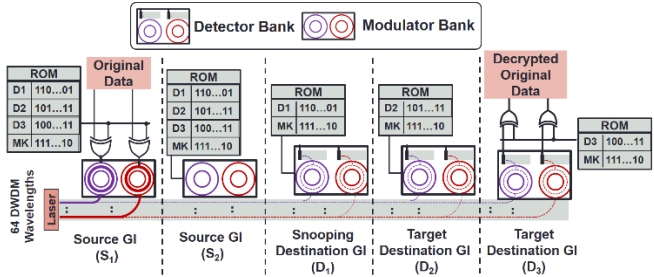

Fig. 5: Overview of proposed PV-based security-enhancing Privy Data Encipherment Scheme (*PDES*).

**Unicast communication example:** Suppose $S_1$ wants to send a data packet to $D_3$. $S_1$ first accesses the 512-bit *unicast key* corresponding to $D_3$ from its local ROM and XORs the data packet with this key, and then transmits the encrypted data packet over the link. At $D_3$, the data packet needs to be received and then decrypted using the correct key (either the *unicast key* or *multicast key*) that it has. To be able to receive (filter out) the data packet on the waveguide, $D_3$ needs to know that the incoming packet is intended for it. Similarly, to be able to use the correct key for decryption, $D_3$ also needs to know if the received data packet is multicast or unicast. Typically, in PNoCs that use photonic links with multiple destination GIs, the source GI communicates the identity of the target destination GI and type of transmitted data packet beforehand, during the reservation selection phase (SectionVI). Therefore, in our example, $S_1$ would have informed $D_3$ beforehand about the target and type of the transmitted data packet. As a result, $D_3$ is able to receive the packet and select the correct key for decryption. At $D_3$, the received data packet is decrypted by XORing it with the 512-bit *unicast key* corresponding to $D_3$ from its local ROM. In this scheme, even if the malicious destination GI $D_1$ snoops the data intended for $D_3$, it cannot decipher the data as it neither knows the target destination for the snooped data nor can it access the correct key (*unicast key* corresponding to $D_3$) for decryption.

**Multicast communication example:** Suppose $S_1$ wants to multicast a data packet to $D_2$ and $D_3$, then $S_1$ first accesses the 512-bit *multicast key* from its local ROM and XORs the data packet with this key, and then transmits the encrypted data packet over the link. Both $D_2$ and $D_3$ would have been informed about the transmitted packet beforehand, therefore, both $D_2$ and $D_3$ would be able to receive the multicast data packet. The received data packet is then decrypted at both $D_2$ and $D_3$ by XORing it with the 512-bit *multicast key* stored in the local ROMs of $D_2$ and $D_3$. In this scheme, if $D_1$ snoops the multicast data packet, it cannot decipher the data in spite of having access to the correct *multicast key* in its ROM. This is because $D_1$ does not know that its snooped data is multicast, and therefore, it does not know whether to use the *unicast key* or *multicast key* from its ROM for data decryption.

Thus, our *PDES* scheme protects unicast and multicast data communications against snooping attacks in DWDM based PNoCs.

### C. Overheads of Implementing PDES

Here we provide a general discussion of the latency, area, and power overheads of our *PDES* scheme, although the detailed results for the system-level overhead analysis of our *SOTERIA* framework will be discussed later in the evaluation section (Section VIII). As *PDES* generates the *unicast* and *multicast keys* during the testing phase of the CMP chip, the XOR logic used for key generation does not need to be implemented on the CMP chip. As a result, no overhead of key generation is incurred in the CMP chip. Further, the dithering signal based control mechanism used in the key generation process also does not incur any extra overhead, as such a control system is not integrated on the CMP chip for key generation purpose exclusively, rather it is added for in-situ remedying of PV-induced resonance shifts in MRs. Thus, our key generation process using *PDES* does not incur any extra area overhead on the CMP chip. However, the use of the ROM to store the generated keys incurs some area and power overhead at every GI. The resulting overall (system-wide) overhead depends on the underlying PNoC architecture. Moreover, data encryption at the source GI and data decryption at the destination GI, each requires two steps: (i) accessing the key from the ROM, and (ii) XORing the key with the data packet. As the key access step can be overlapped with the reservation selection phase (Section VI) for both data encryption and data decryption, the latency overhead of this step can be ignored. On the other hand, the XORing step incurs 1 extra cycle delay for data encryption and data decryption each. This delay along with the area and power overheads are accounted for in our system-level overhead analysis presented in Section VIII.

### D. Limitations of PDES

This section presents circumstances in which unicast and multicast communications cannot be secured in spite of using *PDES* at the circuit-level, and consequently motivates the need for a complementary security solution at the architecture-level.

**Unicast communications:** For the unicast data protected with *PDES* to be deciphered, a snooping GI must have access to *(i)* the *unicast key* corresponding to the target destination GI, and *(ii)* the identity information of the target destination GI. As discussed earlier, a *unicast key* is stored only at all source GIs and at the corresponding destination GI, which makes it physically inaccessible to a snooping destination GI. However, if more than one GI in a PNoC are compromised due to HTs in their control units and if these HTs launch a coordinated snooping attack, then it may be possible for the snooping GI to access the *unicast key* corresponding to the target destination GI.

For instance, consider the photonic link in Fig. 4. If both $S_1$ and $D_1$ are compromised, then the HT in $S_1$'s control unit can access the *unicast keys* corresponding to $D_1$, $D_2$, and $D_3$ from its ROM and transfer them to a malicious core (a core running a malicious program). Moreover, the HT in $D_1$'s control unit can snoop the data intended for $D_3$ and transfer it to the malicious core. Thus, the malicious core may have access to the snooped data as well as the *unicast keys* stored at the source GIs. Nevertheless, accessing the *unicast keys* stored at the source GIs is not sufficient for the malicious GI (or core) to decipher the snooped data. This is because the compromised ROM typically has multiple *unicast keys* corresponding to multiple destination GIs, and choosing the correct *unicast key* that can decipher data at any given time requires the knowledge of the target destination GI. Thus, *PDES* can secure unicast data communications in PNoCs even if the *unicast keys* are compromised, as long as the malicious GIs (or cores) do not know the target destinations for the snooped data.

**Multicast communications:** Unlike *unicast keys*, *multicast keys* are not secret from snooping destination GIs by design. Rather all destination GIs, including the malicious ones, are expected to have access to the *multicast keys*. Nevertheless, as discussed in Section V.*B*, if a malicious destination GI snoops the multicast data packet, it cannot decipher the data in spite of having access to the correct *multicast key* in its ROM. This is because the malicious GI does not know that its snooped data is multicast, and therefore, it does not know whether to use the *unicast key* or *multicast key* from its ROM for data decryption. Thus, *PDES* can secure multicast data communications in PNoCs, as long as the malicious GIs (or cores) do not know the type of the snooped data.

*In summary*, *PDES* can protect unicast and multicast data from being deciphered by a snooping GI, as long as the communication metadata (i.e., information about the target destination GI and type of data communication) associated with the snooped data can be kept secret from the snooping GI. But unfortunately, many PNoC architectures, e.g., [11], [27], that employ photonic links with multiple destination GIs utilize the same waveguide to transmit both the actual data and communication metadata. In these PNoCs, if a malicious GI manages to tap the communication metadata from the shared waveguide, then it can determine the nature of the communication and its intended target(s), and then access the correct key from the compromised ROM to decipher the snooped data. Thus, there is a need to conceal the communication metadata from malicious GIs (cores). This motivates us to propose an architecture-level solution, as discussed next.

## VI. RESERVATION-ASSISTED METADATA PROTECTION SCHEME

In PNoCs that use photonic links with multiple destination GIs, data is typically transferred in two time-division-multiplexed (TDM) slots called reservation slot and data slot [11], [27]. To minimize photonic hardware, PNoCs use the same waveguide to transfer both slots, as shown in Fig. 6(a). To enable reservation of the waveguide, each destination is assigned a reservation selection wavelength. In Fig. 6(a), $\lambda_1$ and $\lambda_2$ are the reservation selection wavelengths corresponding to destination GIs $D_1$ and $D_2$, respectively. When a destination GI detects only its corresponding reservation selection wavelength in the reservation slot, it knows that the incoming data message in the subsequent data slot will be a unicast. For example, in Fig. 6(a), $D_2$ can detect its corresponding reservation wavelength $\lambda_2$ in the reservation slot to know that data will be unicast to it in the subsequent data slot. Therefore, $D_2$ can switch ON its detector bank for data reception and select the appropriate *unicast key* from its ROM for data decryption. Similarly, a destination GI can know if the incoming data is multicast by detecting multiple reservation selection wavelengths in the reservation slot, informing itself to partially switch ON its detector bank for multicast data reception and select the appropriate *multicast key* for data decryption. Thus, the traditional reservation-assisted method of communication (Fig. 6(a)) utilizes the same waveguide to transmit both the communication metadata (using reservation selection wavelengths) and actual data in two separate TDM slots. This traditional reservation assisted method of communication refers to the MWMR concurrent token stream receiver selection strategy from [35].

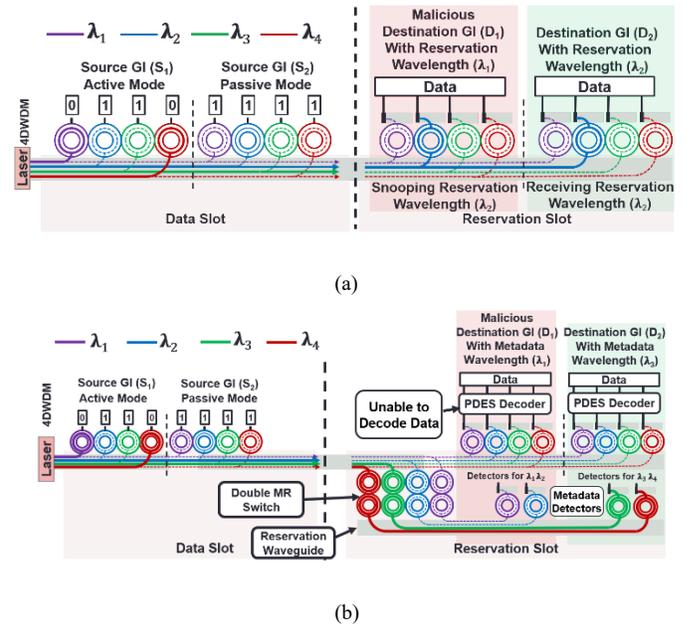

Fig. 6: Reservation-assisted data transmission in DWDM-based photonic waveguides (a) without *RAMPS*, and (b) with *RAMPS*.

However, in the presence of an HT, a malicious GI can tap communication metadata from the shared waveguide during the reservation slot using the same detector bank that is used for data reception. Tapping of communication metadata can provide the malicious GI with important information that enables it to choose the correct encryption key from the compromised ROM to decipher its snooped data. For example, in Fig. 6(a), malicious GI $D_1$ is using one of its detectors to snoop $\lambda_2$ from the reservation slot. By snooping $\lambda_2$, $D_1$ can identify that the data it will snoop in the subsequent data slot will be intended for destination $D_2$. Thus, $D_1$ can now choose the correct encryption key from the compromised ROM to decipher its snooped data.

To address this security risk, we propose an architecture-level Reservation-Assisted Metadata Protection Scheme (*RAMPS*). In *RAMPS*, we add a reservation waveguide, whose main function is to carry reservation slots, whereas the data waveguide carries data slots. This reservation waveguide lays in parallel to the data waveguide, and therefore, it does not intersect with the data waveguide. Rather it is connected to the data waveguide through double MR switches as shown in Fig. 6(b). We use double MRs to switch the signals of reservation slots from the data waveguide to the reservation waveguide, as shown in Fig. 6(b). Double MRs are used instead of single MRs for switching to ensure that the switched signals do not reverse their propagation direction after switching [29]. Compared to single MRs, double MRs also have lower signal loss due to steeper roll-off of their filter responses [29]. The double MRs are switched ON only when the photonic link is in a reservation slot, otherwise they are switched OFF to let the signals of the data slot pass by in the data waveguide.

Furthermore, in *RAMPS*, each destination GI has only two detectors on the reservation waveguide, one of which corresponds to the GI's receiver selection wavelength and the other corresponds to a wavelength signal that transmits communication type information. For example, in Fig. 6(b), $D_1$ and $D_2$ will have detectors corresponding to their reservation selection wavelengths $\lambda_1$ and $\lambda_3$, respectively, on the reservation waveguide. In addition, $D_1$ and $D_2$ will also have detectors corresponding to wavelength signals $\lambda_2$ and $\lambda_4$, respectively, that transmit communication type information. Henceforth, wavelengths $\lambda_1$ ($\lambda_3$) and $\lambda_2$ ($\lambda_4$) are referred to as metadata wavelengths of $D_1$ ($D_2$) and their corresponding detectors on the reservation waveguide are referred to as metadata detectors of $D_1$ ($D_2$). For $D_1$ ($D_2$), the presence of its reservation selection wavelength $\lambda_1$ ($\lambda_3$) and absence of its communication type wavelength $\lambda_2$ ($\lambda_4$) in the reservation waveguide indicates that the incoming data in the next data slot will be unicast to $D_1$ ($D_2$), whereas the presence of both metadata wavelengths $\lambda_1$ ($\lambda_3$) and $\lambda_2$ ($\lambda_4$) indicates that the incoming data will be multicast to multiple destination GIs including $D_1$ ($D_2$). Similarly, the absence of the reservation selection wavelength $\lambda_1$ ($\lambda_3$) indicates that $D_1$ ($D_2$) will not receive any data in the next data slot. Thus, destination GIs can receive important communication metadata on their corresponding metadata wavelengths, using their metadata detectors in the reservation slot. The destination GIs utilize this communication metadata to prepare their detector bank for data reception in the data slot and select the appropriate encryption key (*unicast* or *multicast key*) for data decryption.

The use of the separate reservation waveguide and metadata detectors makes it difficult for the malicious GI $D_1$ to snoop metadata wavelengths ($\lambda_3$ and $\lambda_4$) of $D_2$ from the reservation slot as shown in Fig. 5(b). This is because $D_1$ does not have metadata detectors corresponding to $D_2$'s metadata wavelengths ($\lambda_3$ and $\lambda_4$) on the reservation waveguide. However, the HT in $D_1$'s control unit may still attempt to snoop $D_2$'s metadata wavelengths ($\lambda_3$ and $\lambda_4$) in the reservation slot by retuning $D_1$'s metadata detectors. But succeeding in these attempts would require the HT to perfect the timing and target wavelengths of its snooping attack, which is very difficult due to the large number of utilized metadata wavelengths corresponding to the large number of destination GIs. Thus, $D_1$ cannot know the communication metadata, and therefore, cannot identify the correct key to decipher the snooped data.

*In summary, RAMPS* enhances security in PNoCs by protecting both the actual data as well as the communication metadata from snooping attacks, even if the encryption keys used to secure data are compromised. However, note that the scalability of our RAMPS mechanism is limited by the DWDM capacity of photonic data waveguides in a PNoC. As our RAMPS mechanism requires two dedicated metadata wavelength signals per destination GI, its application is limited to PNoCs that have only up to $N_\lambda/2$ destination GIs connected per data waveguide, given that the waveguide can support multiplexing of only up to $N_\lambda$ wavelength signals.

## VII. IMPLEMENTING SOTERIA ON PNoCs

We characterize the impact of *SOTERIA* on three popular PNoC architectures: Firefly [8], SwiftNoC [35] and LumiNoC [13], all of which use DWDM-based photonic waveguides for data communication. We consider Firefly PNoC with 8×8 SWMR crossbar [8], SwiftNoC PNoC with 32×32 MWMR crossbar [35] with concurrent token stream arbitration, and LumiNoC with 4-layer 1-row photonic network with MWMR waveguides. We adapt the analytical equations from [29] to model the signal power loss and required laser power in the *SOTERIA*-enhanced Firefly, SwiftNoC, and LumiNoC PNoCs. At each source and destination GI of the *SOTERIA*-enhanced Firefly, SwiftNoC, and LumiNoC PNoCs, XOR gates are required to enable parallel encryption and decryption of 512-bit data packets. As discussed in Section V.*C*, we consider a 1 cycle delay overhead for each encryption and decryption of every data packet. The area and energy consumption for 512 XOR gate with 2-bit input is 0.302 nm$^2$ and 0.241 pJ using the 11nm FinFET standard cell library [22]. Moreover, we use nonvolatile ReRAM technology to implement the ROM for key storage, and we model this ReRAM using NVSim [44]. The overall laser power and delay overheads for all of these PNoCs are quantified in the results section.

**Firefly PNoC:** Firefly PNoC [8], for a 256-core system, has 8 clusters (C1-C8) with 32 cores in each cluster. Firefly uses reservation-assisted SWMR data channels in its 8x8 crossbar for inter-cluster communication. Each data channel consists of 8 SWMR waveguides, with 64 DWDM wavelengths in each waveguide. To integrate *SOTERIA* with Firefly PNoC, we added a reservation waveguide to every SWMR channel. Each destination GI has 2 metadata detector MRs on the reservation waveguide where the first one is used to detect the reservation selection wavelength and the second one is used to detect the communication type wavelength. Therefore, in total, this reservation waveguide has 14 metadata detector MRs corresponding to 7 destination GIs. Furthermore, 64 double MRs (corresponding to 64 DWDM wavelengths) are used at each reservation waveguide to implement *RAMPS*. To enable *PDES*, each source GI has a ROM with eight entries of 512 bits each to store seven 512-bit *unicast keys* (corresponding to seven destination GIs) and one 512-bit *multicast key*. In addition, each destination GI requires a ROM with two entries of 512-bits each to store the 512-bit *multicast key* and its own 512-bit *unicast* *key*. In total, *SOTERIA* incurs total photonic hardware overhead of 14 metadata detectors, 64 double MRs, and a reservation waveguide for each SWMR data channel. The area overhead of ROM at a source GI for 8 entries is 0.016µm$^2$ and at a destination GI for two entries is 0.004µm$^2$.

**SwiftNoC PNoC:** We also integrate *SOTERIA* with a 256-core 32-node SwiftNoC PNoC [35] with eight cores in each node and 16 MWMR data channels for inter-node communication. Furthermore, these 256 cores are divided into 4 clusters (each cluster has 64 cores) to enable dynamic re-prioritization and exchange of bandwidth between clusters of cores. Each MWMR data channel has four MWMR waveguides and it connects 32 source GIs and 32 destination GIs. Out of the four MWMR waveguides per MWMR data channel, three waveguides have 64 DWDM wavelengths and one waveguide has 68 DWDM wavelengths. In *SOTERIA*-enhanced SwiftNoC, we add a reservation waveguide to each MWMR data channel. Like Firefly, each destination GI of the SwiftNoC also has 2 metadata detector MRs on each reservation waveguide which are used to detect the reservation selection and communication type wavelengths. Therefore, in total, each reservation waveguide has 64 metadata detector MRs corresponding to 32 destination GIs. To enable *PDES*, each source GI requires a ROM with 32 entries of 512 bits each to store 31 512-bit unicast keys and one 512-bit multicast key, whereas each destination GI requires a ROM with two entries of 512 bits each to store one 512-bit multicast key and its corresponding 512-bit unicast key. SOTERIA incurs total photonic hardware overhead 64 metadata detectors, 64 double MRs, and a reservation waveguide per MWMR data channel. The area overhead of ROM at a source GI for 32 entries is 0.064 µm$^2$ and at a destination GI for two entries is 0.004 µm$^2$.

**LumiNoC PNoC:** Lastly, we integrate *SOTERIA* with a 256-core 64-tile LumiNoC PNoC [13] with 16 MWMR data channels for inter-tile communication. The 64-tiles are arranged in an 8×8 grid with each tile having four cores that are interconnected using a concentrator. Among the 16 MWMR data channels, 8 MWMR channels are laid out horizontally and the remaining 8 MWMR channels are laid out vertically. In the 8×8 grid of tiles, each horizontal MWMR channel connects 8 tiles that constitute one of the 8 rows of the grid, whereas each vertical MWMR channel connects 8 tiles that constitute one of the 8 columns of the grid. Each MWMR data channel in LumiNoC has four MWMR waveguides and connects with total 8 source GIs and 8 destination GIs corresponding to 8 tiles. Each of these four waveguides has 64 DWDM wavelengths which are used for arbitration, receiver selection, and data transfer. In *SOTERIA*-enhanced LumiNoC, we add a reservation waveguide to each MWMR data channel. Similar to Firefly and SwiftNoC, each destination GI of LumiNoC also has 2 metadata detector MRs on the reservation waveguide which are used to detect the reservation selection and communication type wavelengths. Therefore, each reservation waveguide has 16 metadata detector MRs corresponding to 8 destination GIs. To enable *PDES*, each source GI requires a ROM with 7 entries of 512 bits each to store eight 512-bit *unicast keys* and one 512-bit *multicast key*, whereas each destination GI requires a ROM with two entries of 512 bits each to store one 512-bit *multicast key* and one 512-bit *unicast key*. SOTERIA incurs total photonic hardware overhead of 16 metadata detectors, 64 double MRs, and a reservation waveguide per MWMR data channel. The photonic area overhead is evaluated in Section VIII. The area overhead of ROM at a source GI for 8 entries is 0.016 µm$^2$ and at a destination GI for two entries is 0.004 µm$^2$.

**Modeling PV of MR Devices in PNoCs:** Similar to [29] and [57], we adapt the VARIUS tool [20] to model random and systematic die-to-die (D2D) as well as within-die (WID) process variations in MRs for the Firefly, SwiftNoC and LumiNoC PNoCs. We model process variations in MRs in terms of induced resonance shifts in them. The key modeling parameters are mean (µ), variance (σ), and density (ω) of a variable (i.e., MR resonance shift) that follow the normal distribution. The mean (µ) is an MR's nominal resonance wavelength. We consider a DWDM wavelength range in the optical C- and L-bands, with a starting wavelength of 1550 nm and a channel spacing of 0.8 nm, generating a comb of wavelengths that fills the free-spectral-range (FSR). Hence, each modeled MR's mean coincides with the corresponding wavelength in the comb. The variance (σ) of wavelength variation is determined based on the laboratory fabrication data [14] and our target die size. We consider a 256-core chip with a die size of 400 mm$^2$ at a 22-nm process node. For this die size, we consider a WID standard deviation ($σ_{WID}$) in resonance wavelength of 0.61 nm [57] and D2D standard deviation ($σ_{D2D}$) of 1.01 nm [57]. We also consider a density (ω) of 0.5 [20] for this die size, which is the parameter that determines the range of WID spatial correlation required by the VARIUS tool. With these parameters, we use VARIUS to generate 100 PV maps of the 400 mm$^2$ chip. Each of these maps contains over 1 million parts, each of which has an associated value that indicates what the resonance shift in an MR would be if the MR is implemented on that part of the chip. We map the schematic physical layouts of our considered PNoC architectures on to these PV maps to estimate the locations of MRs in the PNoCs on the PV maps, to consequently determine the PV-induced resonance shifts in the MRs.

## VIII. EVALUATION

### A. Evaluation Setup

To evaluate our proposed *SOTERIA* (*PDES*+*RAMPS*) security enhancement framework, we integrate it with the Firefly [8], SwiftNoC [35], and LumiNoC [10] PNoCs, as explained in Section VII. We modeled and performed simulation based analysis of the *PDES*-enhanced and *SOTERIA*-enhanced Firefly, SwiftNoC, and LumiNoC PNoCs using a cycle-accurate SystemC based NoC simulator, for a 256-core single-chip architecture at 22nm. Microarchitectural parameters of the manycore system are presented in Table I. We validated the simulator in terms of power dissipation and energy consumption based on results obtained from the DSENT tool [22]. We used real-world traffic from the PARSEC benchmark suite [23]. GEM5 full-system simulation [24] of parallelized PARSEC applications was used to generate traces that were fed into our NoC simulator. We set a "warmup" period of 100 million instructions and then captured traces for the subsequent 1 billion instructions. These traces are extracted from parallel regions of execution of PARSEC applications. The applications considered for our analysis are stream clusters (SC), bodytrack (BT), canneal (CN), facesim (FS), blackscholes (BS), ferret (FT), swap-tions (SW), fluidanimate (FA), vips (VI), dedup (DD), freqmine (FQ), and X-264. We performed geometric calculations for a 20mm×20mm chip size, to determine lengths of SWMR and MWMR waveguides in Firefly, SwiftNoC, and LumiNoC. Based on this analysis, we estimated the time needed for light to travel from the first to the last node as 8 cycles at 5 GHz clock frequency [35] for Firefly and SwiftNoC PNoCs. Furthermore, we also estimated that the time needed for light to travel from the first to the last node of LumiNoC's MWMR waveguide is 4 cycles at the same clock frequency. We use a 512-bit packet size, as advocated in the Firefly, SwiftNoC and LumiNoC PNoCs.

Table I, Micro-Architectural Parameters for Manycore System

| Number of cores | 64 |
|---|---|
| Threads per core | 1 |
| Per Core: | |
| L1 I-Cache size/Associativity | 32 KB/Direct Mapped Cache |
| L1 D-Cache size/Associativity | 32 KB/Direct Mapped Cache |
| L2 Coherence | MOESI |
| Frequency | 2 GHz |
| Issue Policy | In-order |
| Memory controllers | 8 |
| Main memory | 8 GB; DDR5@30 ns |

The static and dynamic energy consumption values for electrical routers and concentrators in PNoCs are based on results from DSENT [22]. We model and consider the area, power, and performance overheads for our framework implemented with the PNoCs as follows. *SOTERIA* with Firefly, SwiftNoC, and LumiNoC PNoCs has an electrical area overhead of 12.7mm$^2$, 3.4mm$^2$, and, 6.5mm$^2$, respectively, and power overhead of 0.44W, 0.36W, and, 0.42W, respectively, using gate-level analysis and the CACTI 7.0 [25] tool for memory and buffers. The photonic area overhead of implementing our SOTERIA on Firefly, SwiftNoC, and LumiNoC PNoCs is 0.55mm$^2$, 1.18mm$^2$, and 0.41mm$^2$ respectively, based on the physical dimensions [21] of their waveguides, MRs, and splitters. These area overheads are expected to increase the total photonic area in Firefly, SiwftNoC, and LumiNoC PNoCs by 8.11%, 2.36%, and 3.14%, respectively. For energy consumption of photonic devices, we adapt model parameters from recent work [26], [28] with 0.42pJ/bit for every modulation and detection event and 0.18pJ/bit for the tuning circuits of modulators and photodetectors. Photonic Power loss and Crosstalk coefficients are shown in Table II.

Table II, Photonic Power Loss and Crosstalk Coefficients from [29] and [49].

| Parameter type | Parameter value |
|---|---|
| Propagation loss | - 0.274 dB per cm |
| Bending loss | -0.0085 dB per 90° |
| power splitter loss | -0.2 dB |
| MR Q-factor | 9000 |
| MR radius | 5µm |
| Detector responsivity | 0.8 A/W |
| Laser wall plug efficiency | 10% |

The MR current injection tuning power is 130µW/nm from [54] for MRs with 3µm radius. However, from [59], tuning power and efficiency depend on MR size, and from [43] the MRs in this work have to be tuned for up to half the channel gap only. Therefore, as we consider the channel gap of 0.8nm, the value of 130µW/nm from [54] becomes 86.4µW per MR (≈130µW/nm×(5µm/3µm)×(0.5×0.8nm)) for current injection based tuning power (to remedy PV-induced red shifts). Similarly, the tuning power for heating (to remedy PV-induced blue shifts) of ~650µW/nm from [53] (i.e., up to 8.2nm tuning range at 5.4mW heater power) becomes ~260µW per MR (≈650µW×(0.5×0.8nm)) in this work. These values are listed in Table III, along with other power values.

Table III, Power or energy-per-bit (EPB) values for various components of E/O and O/E conversion and resonance control modules.

| Component | EPB/Power |
|---|---|
| E/O and O/E Conversion | |
| SerDes [51] | 0.5 pJ/bit |
| Receiver [52] | 0.075 pJ/bit |
| Modulator Driver [52] | 0.154 pJ/bit |
| Resonance Control | |
| Dithering Signal Based Control Circuits [15] | 385 µW/MR |
| Thermal Tuning [53] | 260 µW/MR |
| Carrier Injection Tuning [54][59] | 86.4 µW/MR |

### B. Overhead Analysis of SOTERIA on PNoCs

Our first set of experiments compares the baseline (without any security enhancements) Firefly (FR), SwiftNoC (SW), and LumiNoC (LU) PNoCs with their *PDES* and *SOTERIA* enhanced variants. From Section VII, 8 SWMR data channels of the Firefly, 16 MWMR data channels of SwiftNoC, and 16 MWMR data channels of LumiNoC are equipped with *PDES* encryption/decryption as well as reservation waveguides for the *RAMPS* scheme.

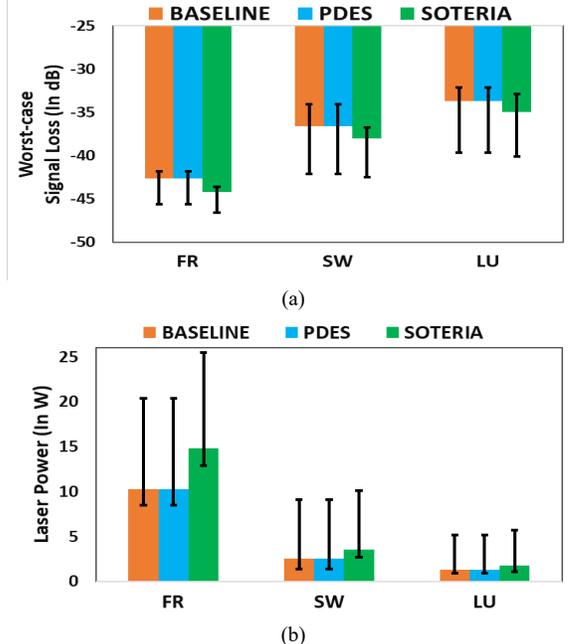

Fig. 7: Comparison of (a) worst-case signal loss (in dB) and (b) laser power dissipation of *SOTERIA* framework on Firefly, SwiftNoC, and LumiNoC PNoCs with their respective baselines considering 100 process variation maps.

We adapt the analytical models from [29] to calculate the total signal loss at the detectors of the worst-case power loss node ($N_{WCPL}$), which corresponds to router C4R0 for the Firefly PNoC [8], node $R_{63}$ for the SwiftNoC PNoC [35], and tile $T_{56}$ for the LumiNoC PNoC [13]. Fig. 7(a) summarizes the worst-case signal loss results for the baseline and *SOTERIA* configurations for the three PNoC architectures. From the figure, we can observe that all the PNoC architectures with *PDES* has no extra worst-case signal loss compared to their respective baselines. *PDES* encoding scheme do not add additional hardware on the photonic waveguide, therefore no additional losses are incurred in PNoC variants with *PDES*. Furthermore, Firefly PNoC with *SOTERIA* increases loss by 1.7dB, SwiftNoC PNoC with *SOTERIA* increases loss by 1.39dB, and LumiNoC PNoC with *SOTERIA* increases loss by 1.3dB, on average, compared to their respective baselines. Compared to the baseline PNoCs that have no single or double MRs to switch the signals of the reservation slots, the double MRs used in the RAMPS scheme of *SOTERIA*-enhanced PNoCs to switch the wavelength signals of the reservation slots increase through losses in the waveguides, which ultimately increases the worst-case signal losses in the *SOTERIA*-enhanced PNoCs. Using the worst-case signal losses shown in Fig. 7(a), we determine the total photonic laser power and corresponding electrical laser power (using laser wall-plug efficiency of 3% [28]) for the baseline and *SOTERIA*-enhanced variants of Firefly, SwiftNoC, and, LumiNoC PNoCs, shown in Fig. 7(b). From this figure, the Firefly, SwiftNoC, and LumiNoC PNoCs with *SOTERIA* have laser power overheads of 44.6% 37.7%, and 35% on average, compared to their baselines.

subsection. From Fig. 7(a), Firefly with *PDES* and *SOTERIA* has 5.2%, SwiftNoC with *PDES* and *SOTERIA* has 9.5%, and, LumiNoC with *PDES* and *SOTERIA* has 14.2% higher latency on average compared to their respective baselines. The additional delay due to encryption and decryption of data (Section VII.A) with *PDES* contributes to the increase in average latency. In addition, *RAMPS* scheme in *SOTERIA* does not contribute any additional cycles to packet transfer, therefore this plot shows no increase in average latency across all the PNoCs between *PDES* and SOTERIA variants. Furthermore, from this plot it can be observed that the average packet latency of SwiftNoC PNoC is higher compared to Firefly PNoC. This is because in SwiftNoC, all data packets traverse through photonic MWMR channels, and therefore they all require encryption-decryption that incurs a latency overhead. On the other hand, in Firefly which is a hybrid electric-photonic architecture, only a few data packets that traverse through the photonic SWMR channels require encryption-decryption, and therefore, only a few data packets incur the related latency overhead, reducing the average packet latency for Firefly compared to SwiftNoC. Moreover, the average packet latency for LumiNoC is higher compared to the other PNoCs, because a significant number of data packets in this PNoC switch between horizontal and vertical MWMR channels, and hence, undergo the encryption-decryption process twice, which in turn increases average packet latency for LumiNoC PNoC.

From the results for EDP shown in Fig. 8(b), Firefly with *PDES* has 2.9%, SwiftNoC with *PDES* has 7.4%, and LumiNoC with *SOTERIA* has 11.8% higher EDP on average compared to their respective baselines. Increase in EDP for the *PDES*-enhanced PNoCs is mainly due to the increase in their average packet latency. In addition, a dynamic and static energy consumption in encryption and decryption circuitry of *PDES* also contributes to increase in EDP. Furthermore, from this plot it can also be seen that Firefly with *SOTERIA* has 5.3%, SwiftNoC with *SOTERIA* has 11.3%, and LumiNoC with *SOTERIA* has 14.6% higher EDP on average compared to their respective baselines. Increase in EDP for the *SOTERIA*-enhanced PNoCs is not only due to the increase in their average packet latency contributed by *PDES*, but also due to the presence of the additional *RAMPS* reservation waveguides, which increases the required photonic hardware (e.g., more number of MRs) in the *SOTERIA*-enhanced PNoCs. This in turn increases static energy consumption (i.e., laser energy and tuning energy), ultimately increasing the EDP. From the results presented in this section, we can conclude that the *SOTERIA* framework improves hardware security in PNoCs at the cost of additional laser power, average latency, and EDP overheads.

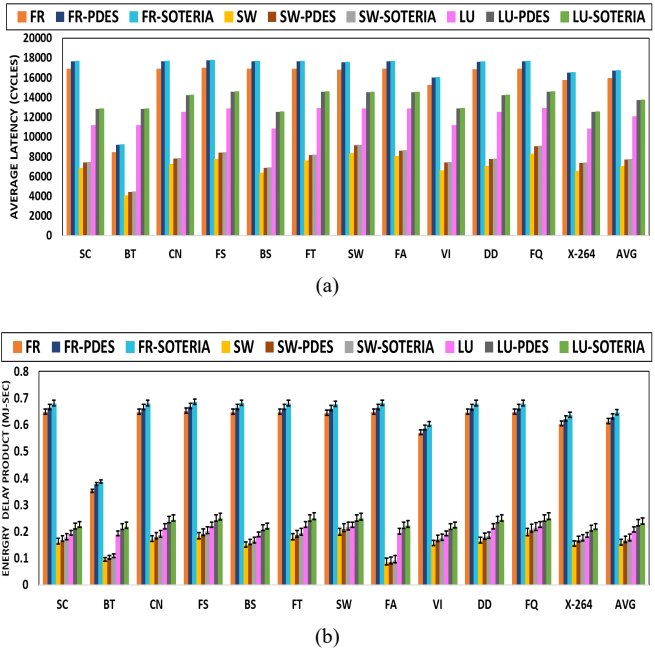

Fig. 8: (a) Normalized average latency and (b) energy-delay product (EDP) comparison between different variants of Firefly, SwiftNoC, and LumiNoC PNoCs for PARSEC benchmarks. Bars represent mean values of EDP for 100 PV maps; confidence intervals show variation in EDP across 100 PV maps. The latency results do not get affected by PV, therefore the average latency bars do not have confidence intervals.

Fig. 8 presents detailed simulation results that quantify the average packet latency and energy-delay product (EDP) for the three configurations (i.e., baseline, *PDES*) of the Firefly, SwiftNoC, and LumiNoC PNoCs. Results are shown for twelve multi-threaded PARSEC benchmarks discussed in the previous

### C. Analysis of Overhead Sensitivity

Our last set of evaluations explore how the overhead of *SOTERIA* changes with varying levels of security in the network. Typically, in a CMP, only a certain portion of data that contains sensitive information (i.e., keys) and only a certain number of communication links may need to be secured. Therefore, for our analysis in this section, instead of securing all data channels of the SwiftNoC (SW) PNoC, we secure only a certain number channels using *SOTERIA*. Out of the total 16 MWMR channels in the SwiftNoC PNoC, we secure 2 (FX-SOTERIA-2), 4 (FX-SOTERIA-4), 8 (FX-SOTERIA-8), and 12 (SW-SOTERIA-12) channels, and evaluate the average packet latency and EDP for these variants of the *SOTERIA*-enhanced SwiftNoC PNoC.

In Fig. 9, we present average packet latency and EDP values for the five *SOTERIA*-enhanced configurations of SwiftNoC. From Fig. 9(a), SW-SOTERIA-2, SW-SOTERIA-4, SW-SOTERIA-8, and, SW-SOTERIA-12 have 1.2%, 2.5%, 4.8%, and 7.4% higher latency on average compared to the baseline SwiftNoC. An increase in the number of *SOTERIA* enhanced MWMR waveguides increases the number of packets that are transferred through the *PDES* scheme, which contributes to the increase in average packet latency across these variants. From the results for EDP shown in Fig. 9(b), SW-SOTERIA-2, SW-SOTERIA-4, SW-SOTERIA-8, and, SW-SOTERIA-12 have 1.3%, 2.7%, 5.3%, and 8% higher EDP on average compared to the baseline SwiftNoC. The EDP in SwiftNoC increases with an increase in the number of *SOTERIA* enhanced MWMR waveguides. In addition to the increase in average packet latency, the increase in signal loss due to the higher number of reservation waveguides, double MRs for switching, and metadata detector MRs increases overall power, and thus EDP across these variants.

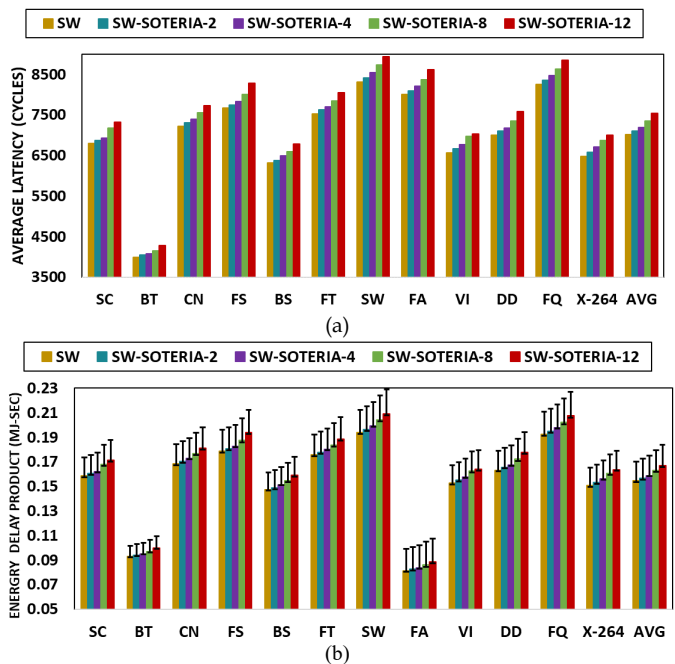

Fig. 9: (a) Normalized latency and (b) energy-delay product (EDP) comparison between SwiftNoC baseline and SwiftNoC with 2, 4, 8, and 12 SOTERIA enhanced MWMR data channels, for PARSEC benchmarks. Latency results are normalized to the baseline SwiftNoC results.

## IX. CONCLUSION

We presented a novel security enhancement framework called SOTERIA that secures data during unicast and multicast communications in DWDM-based PNoC architectures from snooping attacks. The proposed SOTERIA framework shows interesting trade-offs between security, performance, and energy overheads for the Firefly, SwiftNoC, and LumiNoC PNoC architectures. Our analysis shows that SOTERIA enables hardware security in crossbar based PNoCs with minimal overheads of up to 14.2% (as low as 1.2%) in average latency and of up to 14.6% (as low as 1.3%) in EDP compared to the baseline insecure PNoCs. Thus, SOTERIA represents an attractive, low-overhead solution to enhance hardware security in emerging DWDM-based PNoCs.

**Sai Vineel Reddy Chittamuru** (S'14, M'18) is currently with the Micron Technology, Inc, Austin, TX, USA. He received his Ph.D. degree in Electrical Engineering from Colorado State University, Fort Collins, CO, USA in 2018. His research interests include embedded system, systems-on-chip and optical networks-on-chip. Dr. Chittamuru is the recipient of the best paper awards from IEEE/ACM SLIP 2016 and ACM GLSVLSI 2015 conferences, and a best paper nomination from the IEEE ISQED 2016 conference for his research contributions.

**Ishan Thakkar** (S'14, M'18) received his Ph.D. degree in Electrical Engineering from Colorado State University, Fort Collins, CO, USA in 2018. He is currently an Assistant Professor of Electrical and Computer Engineering at University of Kentucky, Lexington, KY, USA. His research interests include photonic networks-on-chip, high-speed optical interfaces, and manycore hardware security. Dr. Thakkar is the recipient of the best paper award from IEEE/ACM SLIP 2016 conference and a best paper nomination from the IEEE ISQED 2016 conference for his research contributions.

**Sudeep Pasricha** (M'02, SM'13) received his Ph.D. degree in computer science from the University of California, Irvine in 2008. He is currently a Professor of Electrical and Computer Engineering at Colorado State University. His research interests include energy efficiency and fault tolerance for multicore computing. Dr. Pasricha is in the editorial board of various journals such as TCAD, TECS, JETC, etc. He currently is or has been in the Organizing Committee and TPC of various conferences such as DAC, DATE, NOCS, ESWEEK, VLSID, etc. He has received Best Paper Awards at the GLSVLSI'18, NOCS'18, SLIP'16, GLSVLSI'15, AICCSA'11, ISQED'10 and ASPDAC'06 conferences.

**Sairam Sri Vatsavai** pursuing M.S. degree in Electrical Engineering from University of Kentucky, Lexington, USA. He has experience of working as a Software Engineer at EPAM Systems India Pvt Ltd. His research interests include photonics networks-on-chip.

**Varun Bhat** received his M.S. degree in Electrical Engineering from Colorado State University, Fort Collins, CO, USA in 2018. He is currently with Qualcomm, San Diego, CA, USA. His research interests include photonic networks-on-chip and performance simulation.